\documentclass[journal]{IEEEtran}
\usepackage[switch]{lineno} 
\usepackage{lipsum} 

\usepackage{cite}

\ifCLASSINFOpdf
  \usepackage[pdftex]{graphicx}
  \graphicspath{{../Figures/}}
\else
\fi
\usepackage{amsmath}
\usepackage{algorithmic}

\usepackage{array}

\ifCLASSOPTIONcompsoc
  \usepackage[caption=false,font=normalsize,labelfont=sf,textfont=sf]{subfig}
\else
  \usepackage[caption=false,font=footnotesize]{subfig}
\begin{document}

\title{Deep Learning Multi-Horizon Irradiance Nowcasting: A Comparative Evaluation of Three Methods for Leveraging Sky Images}

\title{Deep Learning Multi-Horizon Irradiance Nowcasting: A Comparative Evaluation of Three Methods for Leveraging Sky Images}
%
%

%

\author{
Erling W. Eriksen, Magnus M. Nygård, Niklas Erdmann and Heine N. Riise
\thanks{Erling W. Eriksen, Magnus M. Nygård and Heine N. Riise are with the Institute for Energy Technology. Niklas Erdmann is with the University of Oslo Department of Technology Systems. \textit{(Corresponding author: Erling W. Eriksen)}}%
}

%
%

\maketitle

\begin{abstract}
We investigate three distinct methods of incorporating all-sky imager (ASI) images into deep learning (DL) irradiance nowcasting. The first method relies on a convolutional neural network (CNN) to extract features directly from raw RGB images. The second method uses state-of-the-art algorithms to engineer 2D feature maps informed by domain knowledge, e.g., cloud segmentation, the cloud motion vector, solar position, and cloud base height. These feature maps are then passed to a CNN to extract compound features. The final method relies on aggregating the engineered 2D feature maps into time-series input. Each of the three methods were then used as part of a DL model trained on a high-frequency, 29-day dataset to generate multi-horizon forecasts of global horizontal irradiance up to 15 minutes ahead. The models were then evaluated using root mean squared error and skill score on 7 selected days of data. Aggregated engineered ASI features as model input yielded superior forecasting performance, demonstrating that integration of ASI images into DL nowcasting models is possible without complex spatially-ordered DL-architectures and inputs, underscoring opportunities for alternative image processing methods as well as the potential for improved spatial DL feature processing methods.

\end{abstract}

\begin{table}[h]
\caption{List of acronyms used in the present work}
\label{tab:acronyms}
\centering
\begin{tabular}{ll}
\hline
\textbf{Acronym} & \textbf{Description} \\
\hline
PV   & Photovoltaics \\
GHI  & Global Horizontal Irradiance \\
ASI  & All-Sky Imager \\
FOV  & Field of View \\
NWP  & Numerical Weather Prediction \\
DL   & Deep Learning \\
DNN  & Dense Neural Network \\
CNN  & Convolutional Neural Network \\
LSTM & Long Short-Term Memory \\
CC   & Cloud Classification \\
CS   & Cloud Segmentation \\
CMV  & Cloud Motion Vector \\
CBH  & Cloud Base Height \\
COT  & Cloud Optical Thickness \\
HSI  & Hue Saturation Intensity \\
LOCF & Last Observation Carried Forward \\
PCA  & Principal Component Analysis \\
PI   & Permutation Importance \\
AE   & Absolute Error \\
MAE  & Mean Absolute Error \\
MSE  & Mean Squared Error \\
RMSE & Root Mean Squared Error \\
SS   & Skill Score \\
IFE  & Institute for Energy Technology \\
WMO  & World Meteorological Organization \\
\hline
\end{tabular}
\end{table}

\section{Introduction}

Solar photovoltaics (PV) has more than doubled its share of the renewable energy over the past 5 years, bringing with it the demand for solutions to predict and manage variability in production \cite{IEA_renewables}. Information about the future PV production, given through forecasts, can be used for power plant and grid control, energy markets trading, to minimize imbalance settlements, and reduce forecast error penalties \cite{forecast_for_smartgrid}, \cite{imbalance_trading_KLYVE2023208}, \cite{value_of_forecasts_GANDHI2024113915}. Forecasts can also serve as a tool in systems combining PV  with storage through hydrogen or batteries, where unregulated power generation with short-term variability is likely to result in increased degradation of the system\cite{batterydegradation_solar_7741532}, \cite{PV_batter_mitigation_Alam2014ANA}, \cite{PV_hydrogen_OUABI2024100608}. 

The selection of a forecasting method depends on the target application and required forecast horizon. Numerical Weather Prediction (NWP) and satellite methods are valuable for day-ahead trading, while All Sky Imager (ASI) and satellite methods can be leveraged towards intraday trading and advanced control of grids and power plants.

Forecasts utilizing ASIs usually target horizons 0 to 30 minutes into the future, a horizon range commonly referred to as nowcasting \cite{nowcasting_review}.  In this domain, the state-of-the-art methods consist of Deep Learning (DL) models applying images from ground-based ASIs and historical PV or irradiance time-series data as inputs\cite{nowcasting_review}, \cite{PALETTA_benchmarking}, \cite{nowcasting_benchmarking}. Within DL nowcasting using ASIs, modern approaches vary, where Convolutional Neural Network (CNN) based regression methods are most frequently applied due to their aptitude for image inputs \cite{PALETTA2023100150}. 

Although there have been recent strides, even the most promising nowcasting methods struggle to consistently perform on a level that is useful for applications. This illustrates the significant challenge of predicting the chaotic nature of the atmosphere. All recent development of successful nowcasting methods uses ASI images, making it essential to understand which aspects of ASI images are most useful for DL nowcasting models. Many advances focus on purely data-driven techniques, applying new DL architectures. However, as seen in other AI applications, such as protein folding, including structural and physical information in the DL system can often create a more useful and accurate model \cite{Jumper2021}.

To investigate the contributions of ASI image information for nowcasting, this work compares three different methods of introducing ASI images into a Long Short-term Memory (LSTM) based DL model for solar forecasting. Two of these utilize a CNN module for image inputs or feature maps in a hybrid CNN-LSTM architecture, a promising approach aiming for the separate extraction of useful spatial and temporal patterns \cite{CNNLSTM_atmos14071192}, \cite{CNNLSTM_XU2024120135}. The CNN-LSTM architecture has been explored in earlier works, but a structured mapping of which image qualities or feature extraction schemes are useful for this architecture has not been investigated. Focusing on this, we make the following contributions.
\begin{itemize}

  \item Comparative evaluation of three ways to leverage ASI images for LSTM-based nowcasting models, showing that the performance is highly dependent on the preprocessing of the data. 
  
  \item Demonstration of  dual camera setup for pixelwise cloud segmentation (CS), cloud motion vectors (CMV), solar position, and stereoscopic cloud base height (CBH) as input to a nowcasting DL model
  
  \item High latitude($\approx 60^{\circ}$) evaluation of nowcasting models, expanding the applicability of nowcasting northward.
  
\end{itemize}

These contributions are achieved through the following approach: Firstly, image features are extracted and engineered from an ASI setup. Secondly, an LSTM architecture is constructed, combined with a CNN for image features. Thirdly, this architecture is trained with 29 days of 10 second-resolution,  historical global horizontal irradiance (GHI) data, in combination with either (A) raw RGB images that are interpreted by the CNN and fed into the LSTM, (B) engineered feature maps that are interpreted by the CNN and fed into the LSTM or (C) aggregated features extracted from the feature maps, fed into the LSTM. This results in three different model architectures that are tested on seven days of 10-second-resolution GHI data under various atmospheric conditions. These results are discussed in the context of how ASI images are best processed as inputs to a DL model, how feature extraction and engineering affect model performance, and how irradiance and cloud conditions impact model evaluation. This will inform future ASI nowcasting development and improve nowcasting as a tool for handling the variability associated with solar energy generation.

\section{Methods}

\subsection{Measurement setup and data collection}\label{subsection:measurement_setup}
\begin{figure}[h] 
    \centering
    \includegraphics[width=3.5in]{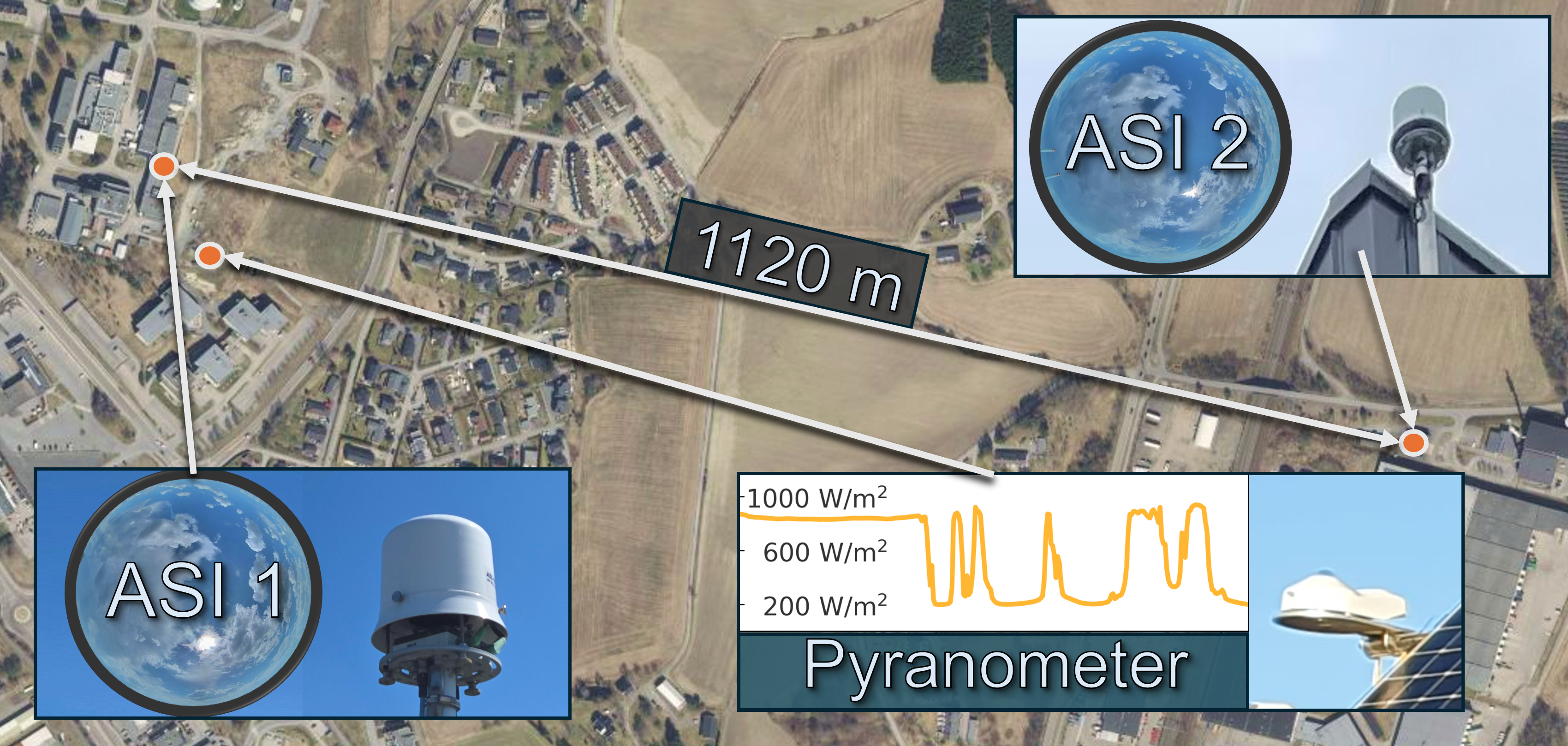} 
    \caption{Map showing the locations of ASI 1 and ASI 2 and the pyranometer measuring GHI. The distance between the two cameras is highlighted, and the inset shows example time series data measured by the pyranometer from 13:30 CET to 14:20 CET on June 15th, 2023.}
    \label{fig:system_map}
\end{figure}

The irradiance data were collected using a spectrally flat ISO9060 Class A Kipp \& Zonen SMP10 pyranometer installed to measure GHI at the Institute for Energy Technology (IFE) in Kjeller, Norway ($59^\circ 58'20.69''$N / $11^\circ 3'8.67''$E) located in a climate classified as \textit{humid continental, warm summer (Dfb)} in the Köppen–Geiger climate classification system. In addition to irradiance measurements, image data was collected from two ASI systems (ASI-16/51 from CMS Ing. Dr. Schreder GmbH), henceforth referred to as ASI 1 and ASI 2. The two-camera setup is a requirement for estimating CBH using stereoscopic methods \cite{NGUYEN2014495}. The position of the ASIs, separated by 1.12 km, and the pyranometer are shown in Fig. \ref{fig:system_map}.  RGB images were captured every ten seconds, with a resolution of 1920$\times$1920 pixels over a period of two years from July 2022 to July 2024. The GHI data was originally recorded with a logging frequency of 1 Hz, but was downsampled by decimation to a frequency of 0.1 Hz to align with the recording frequency of the image data. 

\subsubsection{Image processing}\label{subsection_imageprocessing}
First, to standardize the orientation of ASI 1 and ASI 2 in relation to each other, the images were rotated to align along a North-South axis using the solar position detected in each image for seven clear sky days between February 2023 and September 2023. The sun was found using a linear threshold for the mean intensity of the 8-bit RGB channels of 245, along with a center of mass algorithm implemented in \texttt{SciPy} version 1.11.1. The azimuthal correction was determined to $6.5^{\circ}$ and $ -9.7^{\circ}$ for ASI 1 and ASI 2, respectively. The images were then down-sampled to 100$\times$100 pixels using Lanczos resampling implemented in \texttt{Pillow}, to aid computational efficiency \cite{lanczos_resampling}, \cite{clark2015pillow}. Previous studies report competitive performance down to 64x64 when applying image size reduction with a low-pass filter to avoid unwanted aliasing effects \cite{input_output_SUN}, \cite{PALETTA2023100150}. 

\subsection{Nowcasting methods}\label{subsection:nowcastingmethods}
\begin{figure*}[h]
\centering
\includegraphics[width=5.0in]{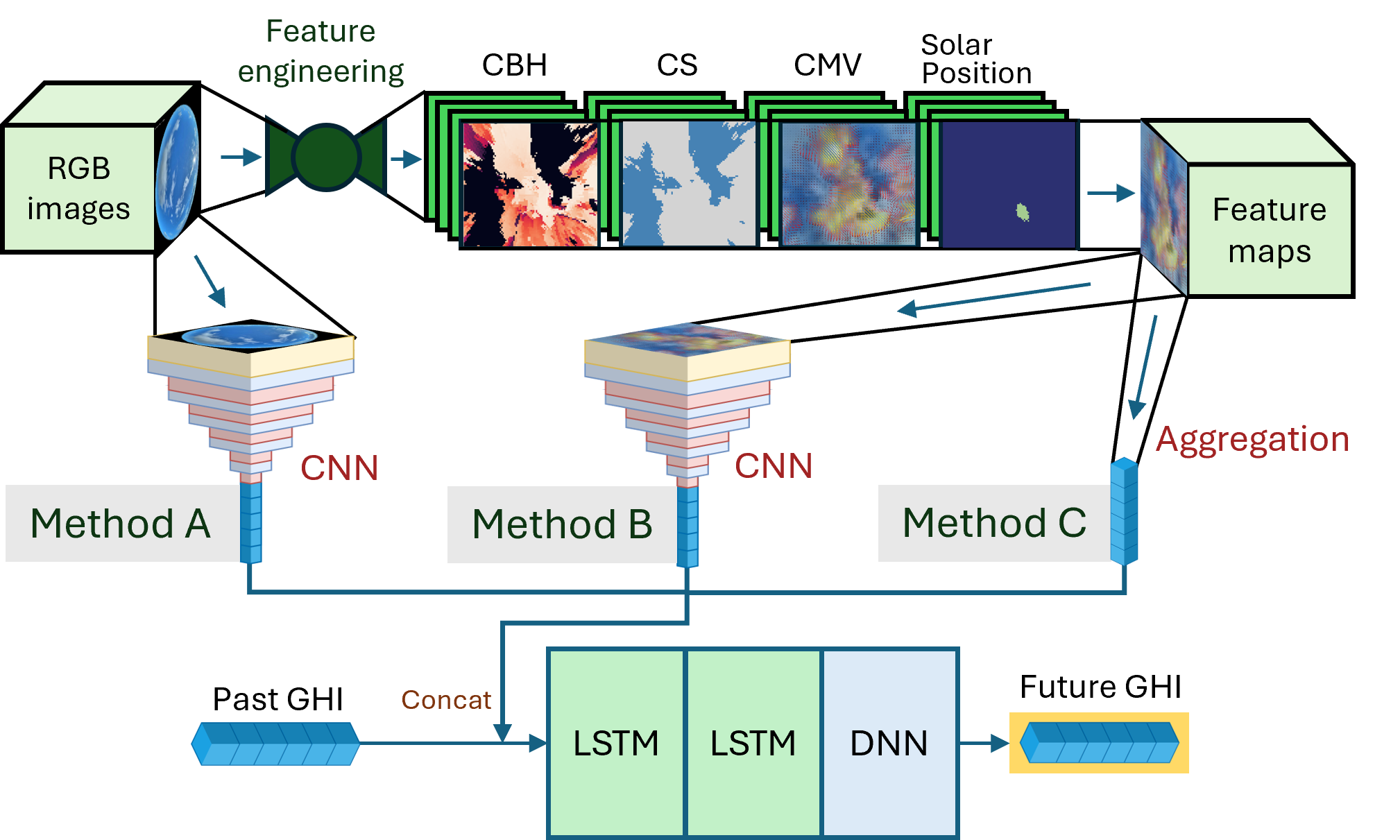}%
\hfil
\caption{Diagram of the three feature engineering and prediction methods A, B, and C. The output from either Method A, B, or C is concatenated with past GHI values as inputs into the LSTM-DNN bulk of the architecture, which is common for all three methods.} 
\label{fig:ABC_scheme}
\end{figure*}

Three methods of leveraging ASI images for DL nowcasting were compared. A diagram of the methods, labeled Method A, Method B, and Method C, is shown in Fig. \ref{fig:ABC_scheme}. 

Three types of neural network structures were applied. The first type, the CNN, has been proven to be suitable for leveraging spatial relationships in spatial data sources\cite{CNN_og_paper}, \cite{AlexNet_NIPS2012_c399862d}. The second type, the LSTM, is a type of recurrent neural network that excels at extracting and using temporal information in sequences\cite{LSTM_hochreiter1997long}. The third is a fully connected Dense Neural Network (DNN) layer with a linear activation, used as a regressor for the final GHI output \cite{DNN}. All methods were implemented using version 2.16.1 of \texttt{TensorFlow} with version 3.3.3 of \texttt{Keras} \cite{tensorflow2015-whitepaper}, \cite{chollet2015keras}.

\subsubsection{\textbf{Method A}} represents an end-to-end DL architecture that takes ASI images along with historical GHI data as inputs. It consists of a CNN constructed to extract and compress information from a series of raw RGB images from ASI 1 into a multivariate input series. The output from the CNN layers, together with historical GHI, are inputted into a 2-layer LSTM network that extracts temporal information. A 2-layer LSTM was chosen due to its ability to handle complex temporal patterns over different timescales, which gave increased performance in initial testing \cite{DEEPLSTM_NIPS2013_1ff8a7b5}. The outputs are then passed to a linear DNN layer, which predicts the future GHI values in a multi-horizon forecast. As shown in Fig. \ref{fig:ABC_scheme}, Method A differs from Method B and C by receiving spatially resolved information from the RGB images, but without physically informed engineered features.

\subsubsection{\textbf{Method B}} is similar to Method A, except that instead of downsampled ASI images, it receives engineered feature maps of the same size as the downsampled ASI images as input. These feature maps, described in section \ref{subsection:featureengineering}, include pixelwise CS, CMV, and solar position for ASI 1 and ASI 2, as well as CBH exclusively for ASI 1. The CBH was only extracted for ASI 1 due to the method's high computational cost.  Method B receives both spatial resolution and physical information in the form of maps of the engineered features.

\subsubsection{\textbf{Method C}} utilizes an aggregation detailed in section  \ref{subsection:datasetconstruction} of the engineered feature maps of Method B into multivariate time series, which are then used as input into a 2-layer LSTM network, with a DNN output layer. Since the spatial information from the engineered feature maps is aggregated, this method receives physically informed input without spatial resolution. 

\begin{figure}[h] 
    \centering
    \includegraphics[width=3.5in]{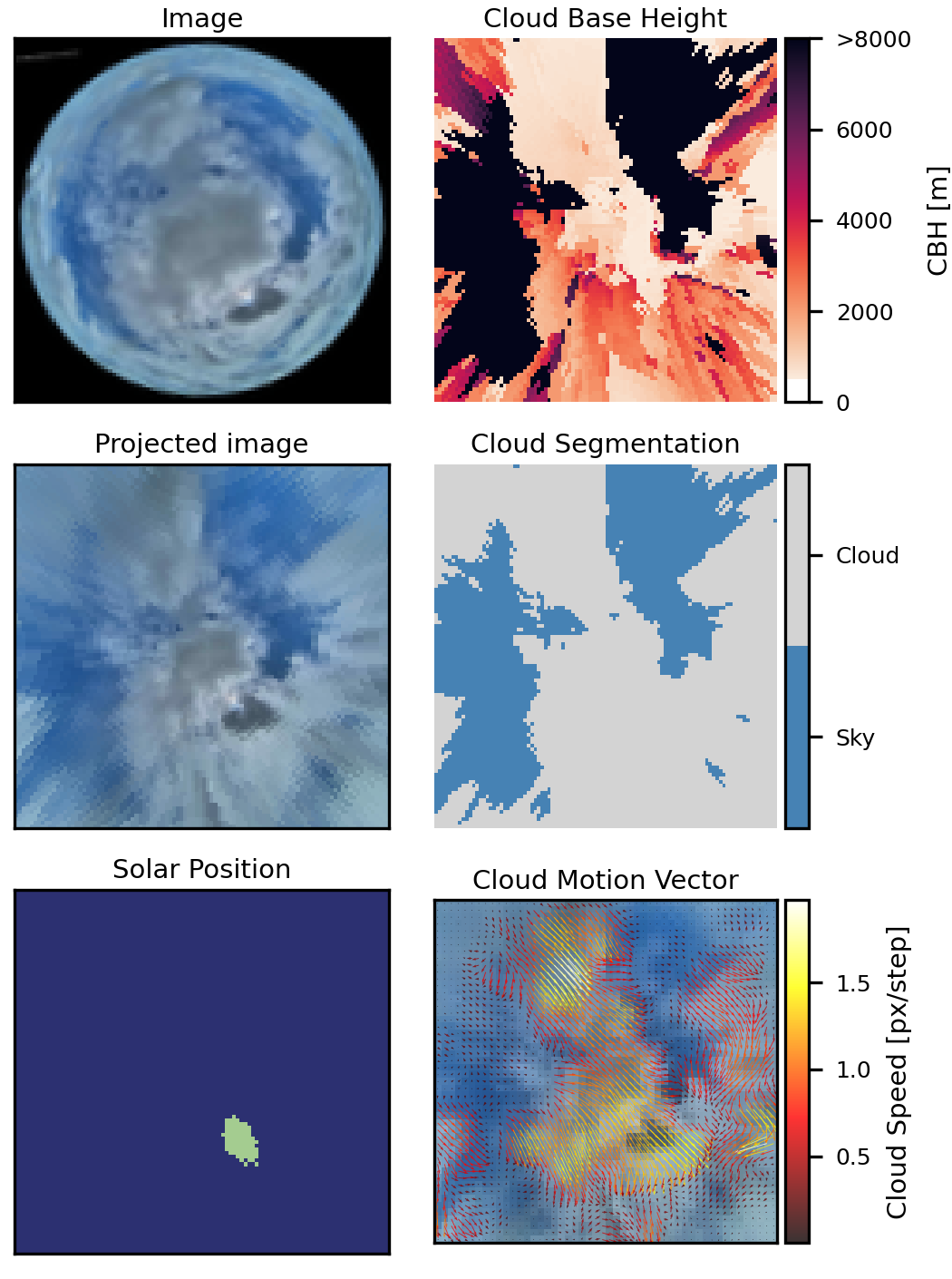} 
    \caption{Example image and feature maps extracted for ASI 1 at 2023-08-04 12:46:10 CET.}
    \label{fig:Feature maps}
\end{figure}  

\subsection{Feature engineering}\label{subsection:featureengineering}
Feature engineering can help compensate for the fact that not all types of information can be synthesized from raw data by a DL model \cite{Feature_Engineering_7506650}, {\cite{feature_engineering_editorial}. To investigate feature engineering of ASI images for DL nowcasting, the following feature engineering techniques were implemented to extract descriptive features from the images captured by the ASIs. 

\subsubsection{Fisheye distortion correction}
To account for the distortion of the zenith in each camera, as caused by the geometry of the fisheye lens, a reverse image projection was used to project the fisheye images to a flat plane. The ASI lens is described by an equidistant projection, characterized by the relation:
\begin{equation}
\label{eq:equidistant_lens}
    \theta_{pixel}/\theta_{FOV} = r_{pixel}/R_{img}
\end{equation} 
where $\theta_{pixel} $ and $ r_{pixel} $ are the zenith angle and image radius of each pixel, respectively. $R_{img}$ is the total image radius and $\theta_{FOV}$ is the zenith of the camera's Field Of View (FOV). Real lens geometries may deviate from the mathematical description, and so a table of deviations from the relationship described in Eq. (\ref{eq:equidistant_lens}) was received from the ASI manufacturer, publicly available in a Zenodo repository \cite{zenodo_rep}. These values were used to create a 7th-order polynomial fit describing the deviations, minimizing the MAE due to distortion correction to under 0.15 $\%$. This polynomial was then used in combination with Eq. (\ref{eq:equidistant_lens}) to translate between the radius of each pixel in the raw image data and its zenith angle. This transformation was also made to subsequent engineered feature map representations. The upper left subpanel in Fig. \ref{fig:Feature maps} shows an example RGB image captured by ASI 1 at 12:46:10 CET on 2023-08-04. The subplot below shows the same image after the described projection and correction. 

\subsubsection{Cloud segmentation}\label{subsection:cloudsegmentation}
The first step in the cloud segmentation of each image was the determination of the sky modality, which describes the number of peaks in the RGB pixel histograms, and is characteristic of the applicability of different types of segmentation thresholds \cite{CS_benchmark_HASENBALG2020596}, \cite{HYTA}. This was done similarly to the method proposed in \cite{HYTA}, using the standard deviation of the saturation (S) in the Hue Saturation Intensity (HSI) image space, defined as:
\begin{equation}
   S = 1- \frac{3}{(R+G+B)}[min(R,G,B)] \label{Saturation}
\end{equation} 
where $R$, $G$ and $B$ are the pixel intensity normalized to a range $[0, 1]$ of the red, green, and blue channels, respectively \cite{rafael_c__gonzalez__2014}. For the camera model and settings, an appropriate threshold between uni- and multimodal images for the standard deviation of $S$ of each image was found through trial-and-error to be $0.09$.

If the image was found to be multimodal, a K-means clustering was performed in the RGB color space to cluster pixels into 3 groups, as shown in Fig. \ref{appendix:fig:cloud_segmentation} in Appendix \ref{appendix:cloud_segmentation} of the Appendix\cite{k_means_macqueen1967some}. K-means was chosen as the segmentation algorithm as it has shown high accuracy in segmenting the image space, without requiring labeled data \cite{d_kmeans_clouds}. Preprocessing of the images with normalization and a dimensionality reduction of the RGB image using principal component analysis (PCA) with 2 components improved speed and separability of the clustering \cite{PCA}. Three clusters of pixel value distributions were identified and classified into clouds, sky, and the frame of the image. If the image excluding the frame was unimodal, meaning overcast or clear weather conditions, a linear threshold was applied on each image using the normalized Blue/Red ratio ($nBR$), as
\begin{equation}
   nBR = \frac{B-R}{B+R}\label{nBR}
\end{equation} 
to detect possible small clouds or gaps in the cloud cover, as proposed by Li et al.\cite{HYTA}.  The thresholds were found through visual inspection to be $0.2$ and $0.35$ for clear and overcast weather conditions, respectively. 

\subsubsection{Cloud Motion Vector} \label{subsection:CMV}
For ASI 1 and ASI 2, the CMV was estimated using subsequent images by application of the Farnebäck Optical flow algorithm implemented in version 4.8.0 of the Python library \texttt{OpenCV} \cite{opencv} \cite{FarnbackFlow}. Using optical flow methods for cloud motion estimation has been found to yield more precise results than window correlation methods \cite{comparison_cloudmotion_ASI_ZAHER20175934}. Farnebäck Optical flow was investigated by Raut et al., comparing extracted cloud motion vectors with radar measurements, concluding long-term stability in the CMV estimates, but with uncertainty likely in part due to measurement and cloud geometry \cite{OpticalFlow_TSI_amt-16-1195-2023}. An example of an extraction can be seen in the bottom right panel of Fig. \ref{fig:Feature maps}. The algorithm was applied using the default parameters, except for a window size set to 4\% of the image size, a polynomial order of 3, a polynomial $\sigma$ of 2.0, and two iterations. Parameters were adjusted through trial and error for a robust CMV extraction without excessive blurring, and to fit with the image format.

The CMV estimation is sensitive to noise from other moving objects in the FOV of the ASI. Close to ASI 2, there is a biofuel heating facility that expels steam from chimneys located in the FOV of the ASI. This affects the calculation of the CMV and cloud segmentation of this camera. In addition, the area is home to birds and insects that occasionally land or ingress onto the pole-mounted ASI, obscuring the view of the sky.  These disturbances were identified and filtered using large CMV inconsistencies between the two cameras. The missing data were then filled by the last observation carried forward (LOCF)  to maintain temporal consistency.  The filters do not remove the impact of these sources of noise completely, but reduce the number of outliers significantly.

\subsubsection{Cloud Base Height} \label{subsection:CBH}
The pixel-wise CBH was estimated through the method proposed by Nguyen and Kleissl \cite{NGUYEN2014495}. Some minor adjustments were made to improve the performance of the algorithm for the available ASI setup. Firstly, to reduce the computational cost, the cloud segmentation was used as a mask to only calculate the CBH for identified cloud pixels. Secondly, the starting correlation window pixel size was reduced to 10$\times$10 to account for lower resolution images and set to a maximum limit of 20\% of the image. Thirdly, when determining the height of each pixel, many pixels in the circumsolar area were mislabeled by the algorithm as having the highest possible height in the search, likely due to the matching of parts of the sun in the correlation windows. To mitigate this, a bin corresponding to an unrealistic cloud height was added, accumulating these pixels along with pixels with a high matching uncertainty, which were both discarded. Finally, if a pixel was discarded but a valid height was found for the same cloudy pixel at the previous timestep, this height was inserted. To ensure the validation made by Kleissl and Nguyen against a ceilometer is representative for the smaller image sizes used in the current work, an investigation of the effect of image size on CBH estimates can be found in Appendix \ref{appendix:CBH_imagesize} of the Appendix.

\subsubsection{Feature dataset construction}\label{subsection:datasetconstruction}
Since Method A takes RGB images as inputs, the RGB images were simply passed as a tensor after downsampling, resulting in tensors of shape $(bs \times lb \times w \times h \times c)$, where $bs$ is the batch size, $lb$ is the lookback, $w$ and $h$ are the the image height and width (in this case 100), and $c$ are the channels corresponding to RGB. Method B takes the 2D feature maps as input, which were stacked in a tensor, resulting in tensors of shape $(bs \times lb \times w \times h \times  c)$, where the number of channels is the number of engineered feature maps from the cameras. Each camera yields solar position, CBH, CS, and two components of cloud motion in the form of the CMV. In addition, both cameras combined yield one CBH mask for ASI 1. Combined, these resulted in a tensor with 9 channels. Method C takes numerical time series as input, which were constructed from the series of each engineered feature map. The CBH was aggregated as the median of the heights of the identified cloudy pixels. The CS was aggregated for each ASI as the ratio of cloudy pixels to the total number of pixels in the image. The CMV was aggregated as the mean independently for the x- and y-directions and for each ASI. The solar zenith and azimuth were used as representations of the solar position. Additionally, the simple extraterrestrial irradiance was used to encode the diurnal pattern of solar irradiation, implemented as
\begin{equation}\label{eq:GHI_clear}
    GHI_{clear}(t) = G_0\times\cos(\theta_z(t)) 
\end{equation} 
where $G_{0}$ is the solar constant, i.e., the extraterrestrial irradiation received by the earth from the sun, and $\theta_z$ is the solar zenith angle. The $GHI_{clear}$ was only used as a feature for Method C to encode the diurnal pattern. This resulted in tensors of shape $(bs \times lb \times  )$, with 11 channels corresponding to the different extracted features.

\subsection{Data, training and evaluation}
\begin{figure}[t!] 
    \centering
    \includegraphics[width=3.5in]{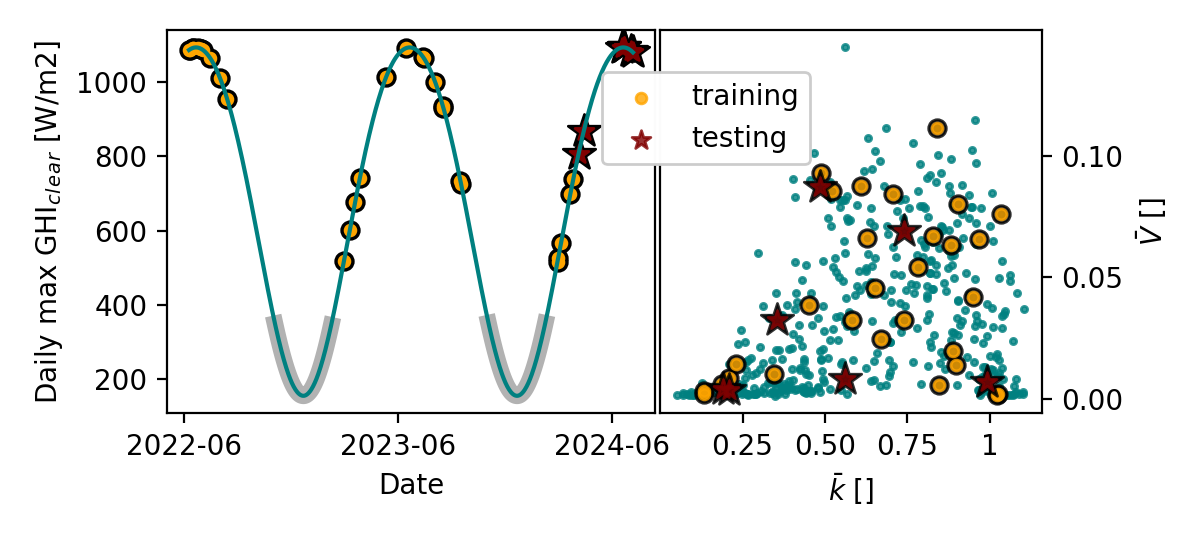} 
    \caption{Left panel shows the distribution of training and testing days with respect to time of year and the maximum GHI$_{clear}$. The gray shaded line indicates days where the maximum solar elevation is below 15$^{\circ}$. The right panel shows the distribution of training and testing days variability and clearness index with respect to 465 other days between 2022-06-09 and 2024-07-07.}
    \label{fig:days_dsitribution}
\end{figure}  

The selected dataset contains 36 days of 10-second-resolution data from 2022-06-09 until 2024-07-07. Timestamps with a solar elevation angle lower than 10$^{\circ}$ were excluded from the dataset, resulting in a total dataset size of 154,817 timestamps. The daily maximum extraterrestrial GHI$_{clear}$, variability index, and clearness index of the days are shown in Fig. \ref{fig:days_dsitribution}. The right panel shows the distribution of days in the training and test set with respect to the daily mean of the clearness index $\bar{k}$ and daily mean variability $\bar{V}$. These are used as proposed in \cite{skillscore}, where
\begin{equation}
\label{eq:daily_clearness}
    \bar{k} = \frac{1}{t_f}\sum_{t_0=0}^{t=t_f}k(t)
\end{equation} 
and 
\begin{equation}
\label{eq:daily_variability}
    \bar{V} = \sqrt{\frac{1}{t_f}\sum_{t_0=1}^{t=t_f}(k(t) - k(t-1))^{2}}
\end{equation} 
where 
\begin{equation}
\label{eq:clearness}
    k(t) = \frac{GHI(t)}{GHI_{clear}(t)}
\end{equation} 
calculated with $GHI_{clear}(t)$ as defined in Eq. \ref{eq:GHI_clear} and measurement $GHI(t)$ from the aforementioned pyranometer. For both $\bar{k}$ and $\bar{V}$, $t_f$ is the last index of the timesteps in a day.

The dataset was split into a training and validation set consisting of 120,339 timestamps in 29 days up until 2024-03-27, while the test set consisted of 33,902 timesteps in 7 days from 2024-04-05 until 2024-07-07, respecting temporal consistency by testing on days in the future with respect to the training data. Out of the training and validation set, 43,381 timesteps were used to validate different model architectures and hyperparameter configurations.

 Fig. \ref{fig:days_dsitribution} shows the distribution of the selected training and testing days within the dataset period with respect to daily maximum GHI, daily mean clearness index $\bar{k}$, and the daily mean variability index $\bar{V}$. The training and test sets are shown within the distribution of 465 days from the same period between 2022-06-09 and 2024-07-07. The figure demonstrates that the selected days in both the training and test sets cover a representative range of the weather conditions experienced at the location. All GHI data were normalized using the solar constant $G_0$.

The hyperparameters of the model were optimized using the training and validation sets. The optimized hyperparameters were batch size, learning rate, training epochs, number of LSTM layers, number of neurons in LSTM layers, LSTM output layer, number of CNN Convolutional (Conv.) layers, number of CNN Conv. filters, CNN Conv. kernel size, and at what layer the CNN output was concatenated to the LSTM layers. Optimization of most of the hyperparameters followed a trial-and-error approach, while the batch size and learning rate were explored using a rough grid search for learning rates ($lr$) from $0.0000005$ to $0.001$, doubling for each step, with a batch size ($bs$) of 1. Then, $lr$ was assumed approximately dependent on $bs$, and scaled by the relation 
\begin{equation}
\label{eq:lr_scaling}
    lr = \frac{New \: bs}{Old \: bs}\times Old\:lr
\end{equation} 
as has been shown empirically in mini-batch scaling of CNNs \cite{lr_bs_scaling}. When approximate $bs$ and $lr$ were determined, a limited search of $[lr/10, lr/5, lr/2, lr, 2lr, 5lr, 10lr]$ was carried out. The number of training epochs was determined using early stopping on the validation set. All models were trained using mean absolute error (MAE) loss and the Adam optimizer algorithm implemented in version 2.16.1 of \texttt{TensorFlow} \cite{kingma2014adam}. The MAE loss was chosen to avoid the strong punishment of the mean-squared error (MSE) loss on large errors, which often leads to an overly smooth forecast. The architectures were kept small, with 42,640 parameters in Method C and 85,300 parameters in Method A and Method B. For a more detailed description of the DL parameters in each of the three methods, see Table \ref{DL_params} in Appendix \ref{appendix:DL_parameters} of the Appendix.

Each model issued multi-horizon forecasts ranging from 10 seconds to 15 minutes, in increments of 10 seconds, totaling 90 individual forecast horizons. Using the RMSE, the MAE, and the skill score $SS$ as
\begin{equation}
    SS = 1 - \frac{RMSE_{forecast}}{RMSE_{reference}}.
    \label{eq:SS}
\end{equation}
The output of each model was evaluated \cite{skillscore}. For the SS, the RMSE$_{reference}$ was the RMSE of a clearness persistence model defined as
\begin{equation}
    f_{persistence}(t+h) = k(t) \times GHI_{clear}(t+h).
    \label{eq:clear_persistence}
\end{equation}
where $h$ is the horizon of the forecast and $GHI_{clear}$ is as defined in Eq. \ref{eq:GHI_clear}. 

\subsection{Feature importance}\label{featureimportance}
To evaluate the importance of the input features, feature permutation importance (PI) was used \cite{feature_importance_RF}. PI shuffles the input dataset of a trained model, one feature at a time. This destroys the temporal information of the feature, while maintaining the feature's exact distribution. The PI is then quantified for each feature as
\begin{equation}
    PI_i  = MAE_{i} - MAE_{ref},
    \label{eq:PI}
\end{equation}
where MAE$_{ref}$ is the MAE for a model predicting on input data without any shuffled features and MAE$_{i}$ is the MAE for a model predicting on input data with feature $i$ shuffled along the time axis. This results in an evaluation of feature usefulness to the model. For each feature $i$, the mean of ten repetitions was used to minimize the effect of stochastic variability in the shuffling.

\section{Results and Discussion}

\subsection{Overall model performance}\label{subsection:modelperformance}
The best models obtained with each of the three methods described in Fig. \ref{fig:ABC_scheme} were evaluated on the test data set. Fig. \ref{fig:ABC_comparison} presents the aggregated performance metrics for Model A, Model B, and Model C for the 7 test days together with the results for the smart persistence model used as reference. The bottom heat map shows the SS, while the top panel shows the RMSE plotted for each horizon. The lower panel demonstrates that all three models outperform the smart persistence model for time horizons longer than 1 minute. On average, Model C performs the best, with an average RMSE of 87.0 W/m$^2$, compared with 87.8 W/m$^2$ for Model A and 90.3 W/m$^2$ for Model B. This is also seen in SS across horizons, where the average SS for horizons over 1 minute for model C is 5.7\%, compared with 5.3 \% for model A and 2 \% for model B. For all three methods, both the RMSE and SS generally increase with horizon, with an exception between 2 and 4 minutes for Models A and B.

\begin{figure}[t] 
    \centering
    \includegraphics[width=3.5in]{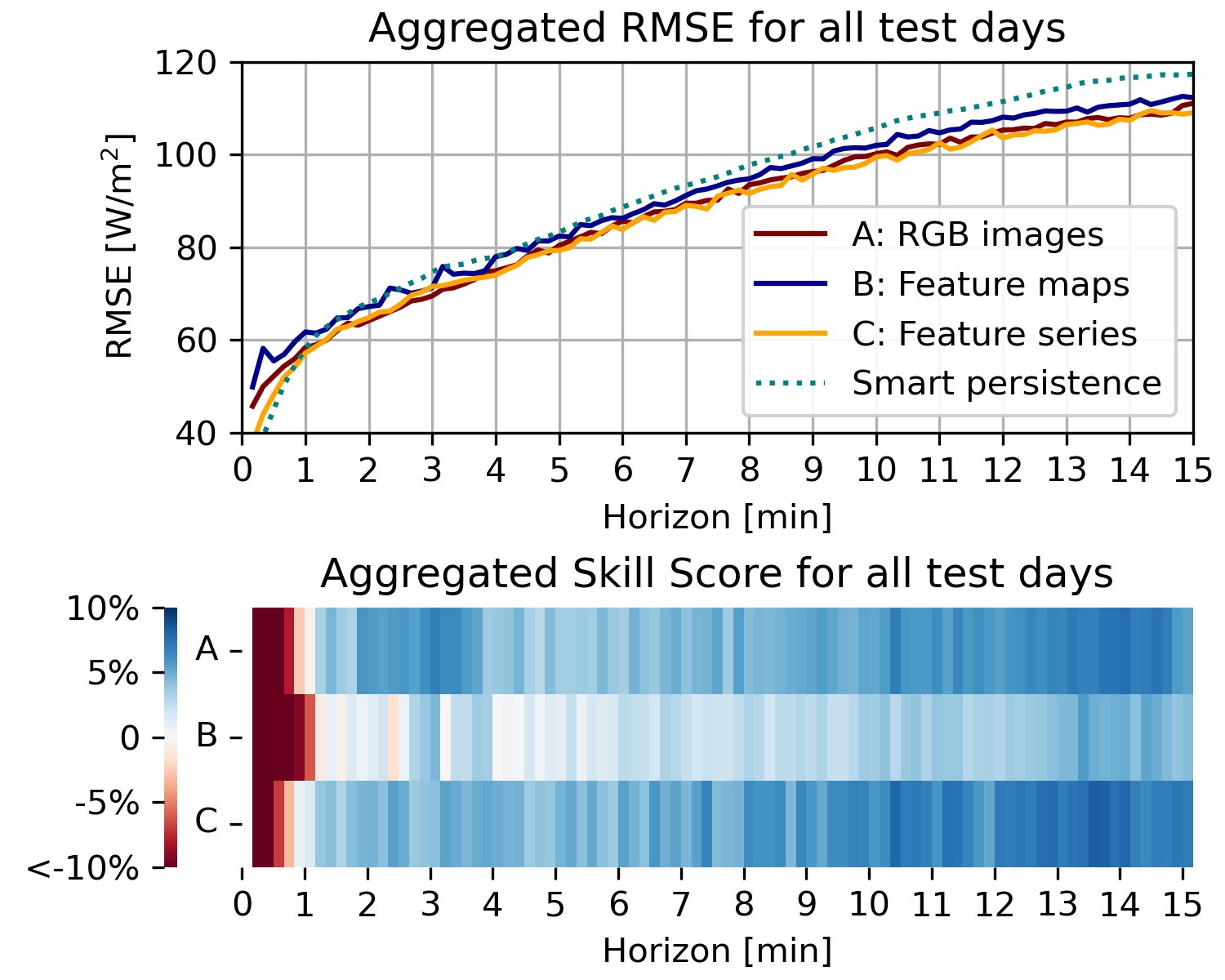} 
    \caption{A comparison of RMSE and SS for models trained using the three methods shown in Fig. \ref{fig:ABC_scheme}.}
    \label{fig:ABC_comparison}
\end{figure}  

\begin{figure*}[h!]
    \centering
    \includegraphics[width=5.0in]{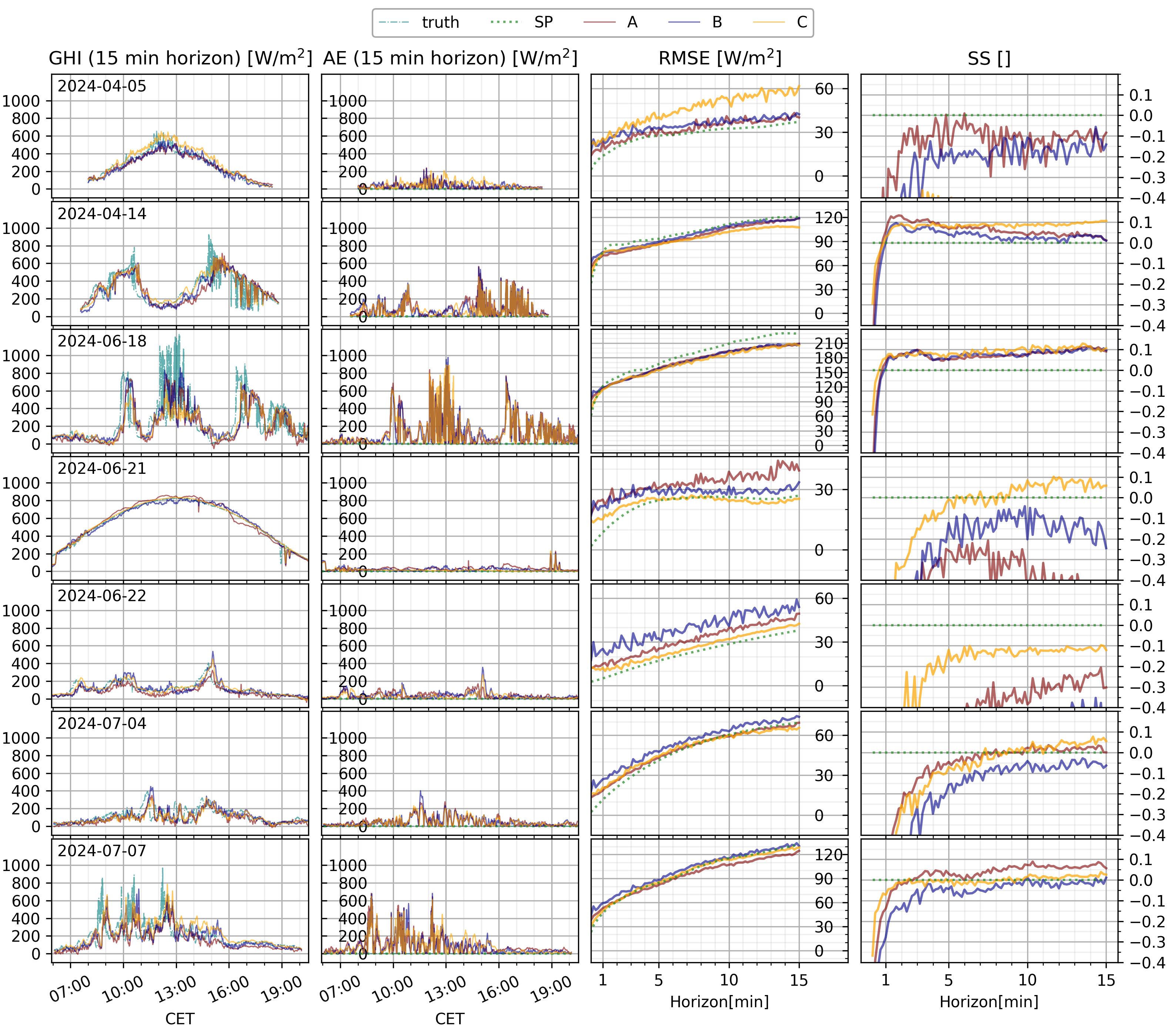}%
    \caption{From the left: The first column shows GHI for each test day, along with forecasts made by Model A, B, and C shown in Fig. \ref{fig:ABC_scheme} for a 15-minute forecast horizon, the second column shows the absolute error (AE) of each forecast made at each timestep. The third column shows the RMSE, while the fourth column shows the SS as defined in Eq. (\ref{eq:SS}). Note that SP refers to the smart persistence model.} 
    \label{fig:forecasts_metrics}
\end{figure*}

\subsection{Daily model performance}\label{subsection:dailyperformance}
Taking a closer look at the metrics for the test dataset, Fig. \ref{fig:forecasts_metrics} shows forecasts and metrics for each of the 7 test days. The first column shows the measured GHI along with the forecasts for a 15-minute horizon, where the second column shows the associated absolute error $AE_i=|y_i-\hat{y}_i|$ for the 15-minute horizon. The third and fourth columns show RMSE and SS, respectively, where the metric is shown for each day and as dependent on forecast horizon.
 
Table \ref{tab:Day_characteristics} shows the daily mean clearness index $\overline{k}$, the daily mean variability index $\overline{V}$, and the cloud classification (CC) based on WMO cloud genera (See Table \ref{CC_genera}) determined through manual inspection using the WMO Cloud Atlas Cloud Identification guide \cite{Cloud_class}, resolved for the morning, midday and evening of the day. The mean daily clearness and variability were calculated from Eq. (\ref{eq:daily_clearness}) and Eq. (\ref{eq:daily_variability}). 

The days in the test set can generally be placed into two categories: Days with a low variability index or days with a high variability index. The days with lower variability are the days 2024-04-05, 2024-06-21, 2024-06-22, and 2024-07-04. These days show low RMSE, ranging from 10 W/m$^2$ to 75 W/m$^2$. Although the RMSE is low, the SS is most often found below 0, showing (as expected) that the performance of a smart persistence model improves with lower variability. For these days, Model C predominantly outperforms Model A and B, except for 2024-04-05, where it underperforms the other models by a large margin. This day is characterized as a completely overcast day with optically thin Altostratus (As) clouds. This causes the irradiance signal to be dampened, but the diurnal trend, i.e., low irradiance in the early morning and late afternoon and higher irradiance around noon, is still present. In these conditions, the cloud optical thickness (COT) is an important quality, as variations within overcast conditions are largely due to variations in COT (accounting for the diurnal cycle). COT is not part of the engineered features of Method B and Method C due to the lack of reliable methods to estimate COT directly from ASI images. Here, Method A has the advantage of flexibility to attempt to extract the most relevant information from the image, regardless of what is deemed physically important or has an available robust method of estimation. Model B has a higher performance under these conditions than Model C, indicating that the richer information present in a spatial representation may compensate to some degree for lacking information in feature representation.

The days in the test set with more variable conditions are 2024-04-14, 2024-06-18, and 2024-07-07. The forecast for these days is characterized by a higher RMSE ranging from 30 W/m$^2$ to 210 W/m$^2$. The skill score is also generally higher than for days with less variability. For these days, Models A and B exhibit similar patterns, where the SS of the models rise and fall for similar horizons, but with a positive offset for Model A, showing a better performance. Model C does not follow the same trends as closely, which is most apparent on 2024-04-14, when the SS is stable across horizons over 1 minute. Model C is better at predicting the last part of the day from 13:00 onward, corresponding to the transition from overcast to partially cloudy skies. Models A and B achieve a higher performance for horizons between 1 and 2 minutes, indicating that the performance degrades as horizons become longer due to an imbalanced focus on transient cloud movements present in the 2D images and features. 

The maximum SS for all methods is also observed on the days with high variability. For Method A, the maximum of 13\%  is found for a horizon of 1 minute and 40 seconds on 2024-04-14, 11\% for Method B for a horizon of 13 minutes and 10 seconds on 2024-06-18, and 13\% for Method C for a horizon of 13 minutes and 20 seconds on 2024-06-18. This shows that the DL methods excel exactly where persistence models fail, when the atmosphere is unstable, and tv he irradiance is likely to change. 

 \begin{table}[!h]
  \renewcommand{\arraystretch}{1.3}
 \caption{Test set daily characteristics: $\overline{k}$, $\overline{V}$, and Cloud Classification (CC) based on the WMO genera (See Table \ref{CC_genera}). Morning is from 6--10, midday from 10--15, evening from 15--20. Slashes between CC genera indicate uncertainty in the determination of CC genera, which were determined using the WMO cloud classification guide \cite{Cloud_class}.}
 \label{tab:Day_characteristics}
 \centering
 \begin{tabular}{|c||c|c|c|}
 \hline
  Day & $\bar{V}$ & $\bar{k}$ & CC    \\
 \hline
  04-05 & 0.0061 & 0.56 & morning: As \\
  & & & midday: As \\
  & & & evening: As \\
 \hline
  04-14 & 0.050 & 0.74 & morning: Cs, Ac\\
  & & & midday: St  \\
  & & & evening: As, Cu into Cl\\
 \hline
  06-18 & 0.061 & 0.48 & morning: Ns  \\
  & & & midday: Ac, Cu, Cb/Ns \\
  & & & evening: Cb/Ns into Cu, Ac \\
 \hline
  06-21 & 0.0039 & 0.99 & morning: Cl \\
  & & & midday: Cl \\
  & & & evening: Cl with small Ac \\
 \hline
  06-22 & 0.0027 & 0.19 & morning: As, Cs \\
  & & & midday: Ns \\
  & & & evening: St/As, Ns \\
 \hline
 07-04 & 0.0023 & 0.20 & morning: Ns \\
  & & & midday: St, Ns/Cb \\
  & & & evening: As/St, Ns \\
 \hline
 07-07 & 0.023 & 0.35 & morning: St, Cu, As  \\
  & & & midday:  Cu, As, Ac, Cb/Ns\\
  & & & evening: Ns, As \\
 \hline
 \end{tabular}
 \end{table}

\begin{table}[!h]
  \renewcommand{\arraystretch}{1.3}
 \caption{WMO cloud genera used in cloud classification \cite{cloudatlas1975}.}
 \label{CC_genera}
 \centering
 \begin{tabular}{|c||c|}
 \hline
  Shorthand/CC & Genera    \\
 \hline
  Cb & Cumulunimbus \\
 \hline
  Ns & Nimbostratus \\
 \hline
  As & Altostratus \\
 \hline
  St & Stratus \\
 \hline
  Cu & Cumulus \\
 \hline
 Ac & Altocumulus \\
 \hline
 Ci & Cirrus \\
 \hline
  Cl & Clear \\
 \hline
 \end{tabular}
 \end{table}

 \begin{figure}[t] 
    \centering
    \includegraphics[width=3.5in]{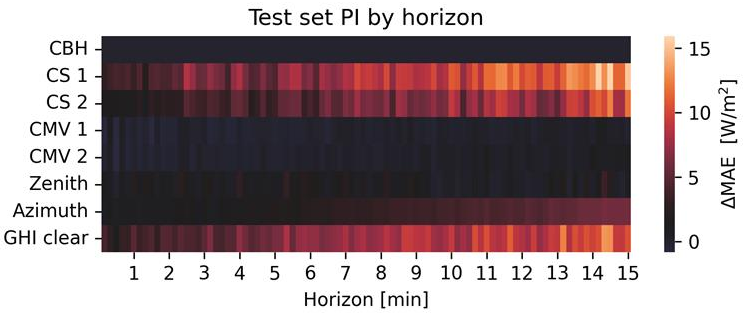} 
    \caption{PI calculated from Eq. \ref{eq:PI} for the features extracted by Method C, aggregated by forecast horizon, where a higher $\Delta$MAE indicates high feature importance. GHI$_{clear}$ is as calculated in Eq. \ref{eq:GHI_clear}. The numbers "1" and "2" refer to the ASI from which the features were extracted (see Fig. \ref{fig:system_map}). }
    \label{fig:C_permutation_importance}
\end{figure}  

\subsection{Feature importance}\label{subsection:Featureimportance}
The ASI features of the best model (Model C) were investigated, with the PI shown in Fig. \ref{fig:C_permutation_importance}, where the upper panel shows the PI aggregated by horizon calculated on the test data. The shuffling of the CS features had the largest impact on the model, along with significant effects from GHI$_{clear}$ and the azimuth. The shuffling of the CBH and CMV features, on the other hand, showed little to no impact on performance. It is also clear that the feature importance increases with an increased horizon, showing that ASI feature inputs become more important as the horizon increases. Supporting evidence of this trend can also be seen in the relationship between the decrease in the autocorrelation of the GHI and the correlations between the other features and the future GHI (See in Fig.\ref{fig:Corellogram} of Appendix \ref{appendix:corellogram} of the Appendix).

The feature importance shows that the model utilizes the CS to a large degree, indicating that improving cloud segmentation algorithms and estimation of cloudiness is a promising route for improving performance. Features that are not used by the model may be discarded to save computation, such as the CBH and CMV. This will reduce the computational time of feature engineering and model training. For the CBH, this also avoids the costly deployment of a second ASI required to implement stereoscopic CBH algorithms. However, each of the methods also contains an inherent uncertainty, created by the aspects of the image data they use (resolution, distortion, noise) and the specifics in their methodologies. As a further step, improving the robustness of these features or enhancing post-processing may make these features also play a role in the performance of future models.

The insight that CS improves performance can also be used in model development, either through the aforementioned improved feature engineering or as an integrated part of the model architecture, for example, as a secondary target, as in \cite{PALETTA2022119924_eclipse}. When developing ASI equipment, one should consider what image qualities may be altered to make sky and cloud more separable, for example, by lowering the camera exposure or using HDR images to minimize overexposure in the circumsolar area and on cloud edges.

\subsection{Method advantages and drawbacks}\label{subsection:methodladvantages}
Determining which of the three approaches is the optimal strategy for the inclusion of ASI images into nowcasting models using DL is a multi-faceted judgment. Each of the methods have advantages and disadvantages related to algorithmic complexity, computational needs, performance and robustness.

For Method A, the main advantage of this end-to-end DL model is the flexibility and ease of using raw RGB data without feature engineering. This avoids the computational expense of some computationally costly calculations, i.e., the pixel-wise stereoscopic CBH, and avoids the inherent uncertainty introduced by the application of feature engineering algorithms. Additionally, this method provides flexibility to the model by allowing useful image qualities to be selected inside the DL method itself. The main advantage of this method is also its drawback. Since the model has to condense the information in the RGB images, as well as extract temporal information, this is a more complex task than using already extracted feature information. This leads to longer training times and requires more training data. This was seen in training, where Method A was trained for 126 epochs, while Model B was trained for 92 epochs. Additionally, since two different ASI images may result in equal engineered features, while different features will always correspond to different ASI images. This means the possible space of features is smaller than the possible configuration of RGB images and the models, therefore, likely requires less data and less training.

The main advantage of Methods B and C is the input of information that is known to be important for the task at hand. Cloud information is important for nowcasting, and so presenting this information can simplify the task, improving performance. In addition, engineered features offer increased explainability, especially when the features are physically informed. By connecting DL features to physical variables of the system, their interpretation also becomes connected to a larger context. The feature importance shows that model C utilizes the CS to a large degree, which can be interpreted as the model values information on how cloudy it is. Such an interpretation is not possible for Method A. The main disadvantage of methods B and C is that a lot of weight is placed on these features being accurate and robust. Additionally, not all information that is known to be important, like the COT, has available robust feature extraction methods that can be used. 

Between Method B and Method C, aggregating physical features to a time-series yields a better performance, requires less storage for input data, and has lower training and inference times. Since the input of Method C is a less resolved version of the input of Method B, it is clear that the CNN in Method B does not yield features more valuable than manual aggregation and post-processing. Method C uses a very simple aggregation technique, the mean, or median, depending on the feature. Spatial information is intuitively valuable for nowcasting, but Model C outperforms Method B, showing that a more complex model, extraction, or aggregation scheme is required to leverage the spatial information of the physically engineered features of Model B.  

A challenge of the performance evaluations in this study, and nowcasting comparative studies in general, is that the day-to-day differences in errors far outscale the differences between the different models. This creates difficulties in comparing the performance of the present models to other literature implementations with other datasets. It also raises the question of how to evaluate forecasts in a way that is useful for PV power plant operators. This is especially challenging in ASI nowcasting, as the high frequency and amount of data make it challenging to include multiple years of data in training and evaluation. With smaller datasets, there is a larger risk of excluding certain relevant atmospheric conditions at a site. Without ensuring evaluation under the same atmospheric conditions, comparisons are, at best, biased and, at worst, misleading. In such a comparison, differences in performance may be due to different atmospheric conditions and not model ability. To tackle this, benchmarking studies and applications on common public datasets are invaluable.

\section{Conclusion}
In this work, three methods of leveraging ASI images for multi-horizon DL nowcasting were compared. A comparison was made between CNN feature extraction from RGB images (Method A), CNN feature extraction from engineered feature maps (Method B), and using engineered features in the form of timeseries (Method C). Each of the feature sets was used as input to LSTM layers predicting GHI 10 s to 15 minutes ahead. 

For the 7-day test dataset consisting of 33,902 timesteps with diverse atmospheric conditions, Method C yielded the highest overall performance. Method C achieved an average RMSE of 87.0 W/m$^2$, compared to 87.8 W / m $^2$ for Method A and 90.3 W / m $^2$ for Method B. Similar results were found for the SS, with some differences depending on the forecast horizon. Analysis of feature importance for method C showed increased reliance on engineered features with increasing forecast horizon. Based on this comparative evaluation, physically informed timeseries as extracted by Method C is recommended for computational parsimony and robust performance when applying LSTM models. Based on the analysis of feature importance, CS, azimuth, and the GHI$_{clear}$ should be provided as features. The exception to this recommendation is in locations with frequent overcast conditions of thin altostratus clouds. This is a limitation of the feature set, which may be amended by future work in robust COT feature extraction.

These results demonstrate the viability of alternative image processing methods for nowcasting, highlighted by the fact that integrating physically relevant timeseries features extracted from ASI images outperformed DL feature extraction using a standard CNN architecture.

\appendices

\section{Example of cloud segmentation}
Fig. \ref{appendix:fig:cloud_segmentation} shows an example of a cloud segmentation, where the two upper plots show the image and segmentation, while the lower plot shows the pixel values in an RGB space.

\begin{figure}[h] 
    \centering
    \includegraphics[width=2.5in]{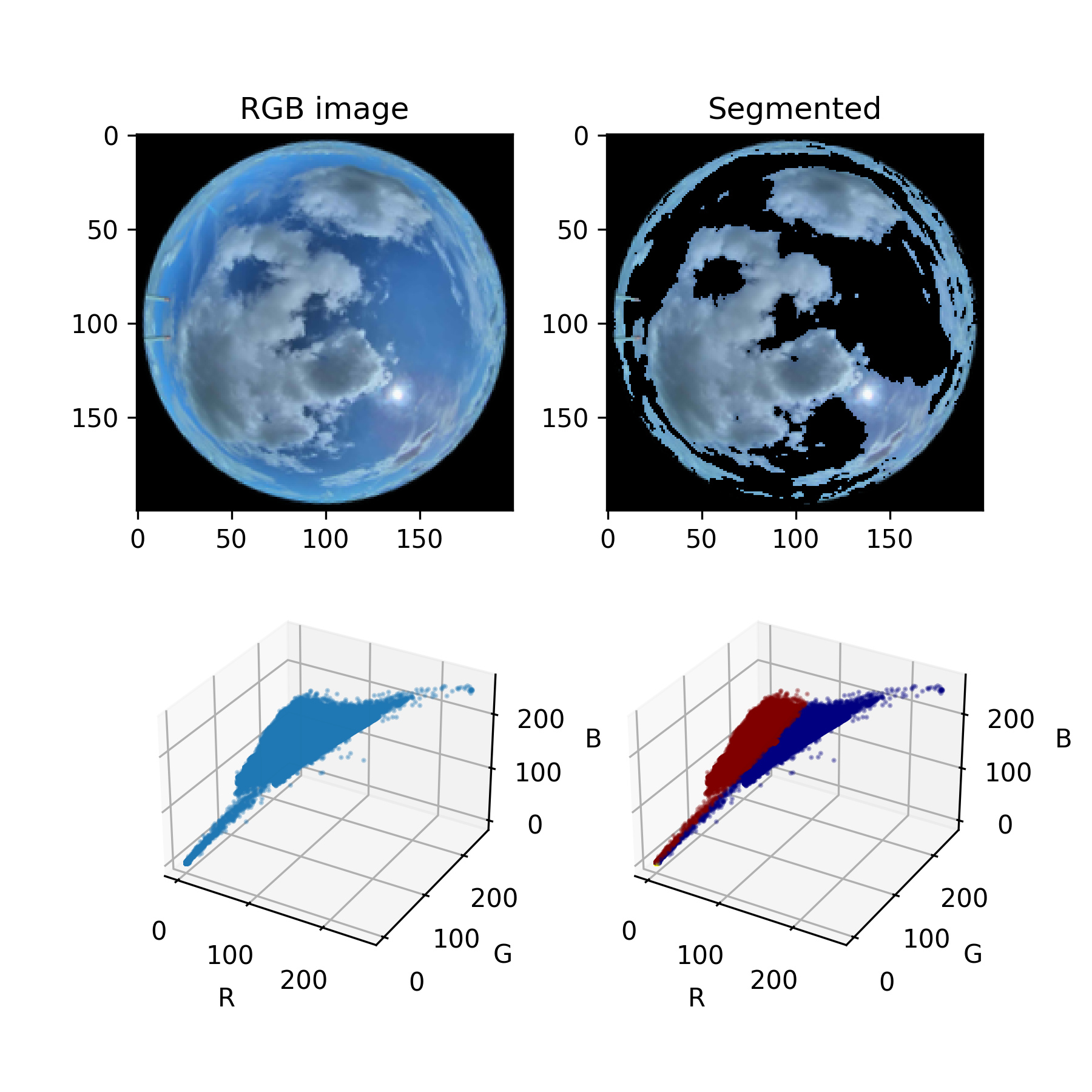} 
    \caption{Example of cloud segmentation using RGB-space k-means clustering.}
    \label{appendix:fig:cloud_segmentation}
\end{figure}  
\label{appendix:cloud_segmentation}

\section{Corellogram ASI features}
\begin{figure}[!h] 
    \centering
    \includegraphics[width=3.5in]{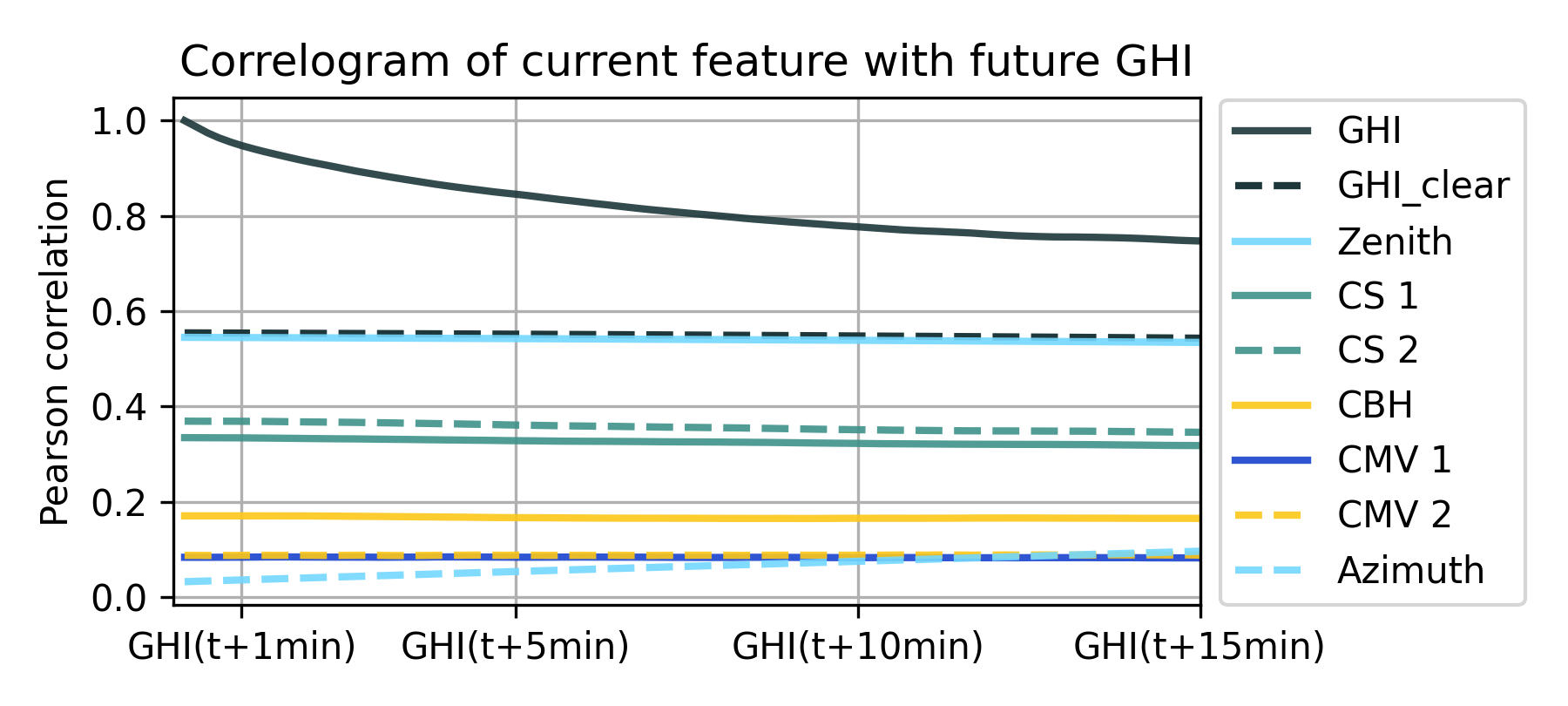} 
    \caption{A corellogram showing the absolute Pearson correlation between the current value of a feature and the future value of the GHI, at a time denoted on the x-axis. Calculated from the 120 339 samples in the training/validation data.}
    \label{fig:Corellogram}
\end{figure}  \label{appendix:corellogram}

\section{DL model parameters for the three methods of the comparison}

 For the DL architectures that were applied, Table \ref{DL_params} shows selected hyperparameters and architecture descriptors of the three methods. 
 The lookback refers to the amount of historical data each method receives. For Methods A and B, this also includes previous RGB images or feature maps, respectively. The horizon is the furthest prediction time of the 90 predictions given for each 10 s intervals. Dropout refers to per layer drop out in the CNN architecture, units refer to the number of units per layer for the described layer type. Convolutional filters are in order from the start to the end of the CNN. Additionally, each convolutional layer is also followed directly by a 2$\times$2 MaxPooling layer. For Methods A and B, in addition to the CNN layers described in Table \ref{DL_params}, a trainable convolutional 1$\times$1 layer was added at the head of the block, functioning as a pixelwise scaling layer. 
 \begin{table}[!h]
 \renewcommand{\arraystretch}{1.2}
 \caption{DL Architecture Parameters for the three methods.}
 \label{DL_params}
 \centering
 \begin{tabular}{|c||c|c|c|c|c|}
 \hline
  Method & A & B & C\\
 \hline
 Input features & RGB & Feature maps &Feature series \\
 \hline
  Parameters & 85,228 & 85,300 & 42,640\\
 \hline
  Lookback & 150 s & 150 s & 150 s\\
  \hline
  Horizon & 15 min & 15 min & 15 min\\
  \hline
  Frequency & 10 s & 10 s & 10 s\\
  \hline
  Conv. layers  & 6 & 6 & -\\
  \hline
  Conv. filters  & 8,8,16,24,32,40 & 8,8,16,24,32,40 & -\\
  \hline
  Conv. kernel size  & 3$\times$3 & 3$\times$3 & -\\
  \hline
  Pooling layers  & 5 & 5 & -\\
  \hline
  Pooling size  & 2$\times$2 & 2$\times$2 & -\\
  \hline
  Padding  & valid & valid & -\\
  \hline
  LSTM layers & 2 & 2 & 2\\
 \hline
  LSTM units & 25,25 & 25,25 & 25,25\\
 \hline
  Dense layers & 1 & 1 & 1\\
 \hline
  Dense units & 90 & 90 & 90\\
 \hline
  Dropout & 0.05 & 0.05 & 0.0\\
 \hline
 Loss & MAE & MAE & MAE\\
 \hline
  Learning rate & $1.5\times10^{-5}$ & $1.5\times10^{-5}$ &$2\times10^{-4}$\\
 \hline
  Batch size & 128 & 128 & 1024\\
 \hline
 Epochs & 126 & 92 & 344\\
 \hline

 \end{tabular}
 \end{table}\label{appendix:DL_parameters}

\section{CBH estimation and image size}\label{appendix:CBH_imagesize}

The stereoscopic algorithm was validated against a ceilometer by the original authors, but for smaller image sizes, as the one used here, its validity is uncertain \cite{NGUYEN2014495}. Fig. \ref{appendix:CBH_resolution} shows the distribution of CBH estimates as a function of changing ASI image resolution. 
Fig. \ref{appendix:CBH_resolution} shows that the estimates are relatively stable down to smaller image sizes, with a small shift towards lower cloud heights. This is likely in part due to the lower resolution decreasing the accuracy of the method, but also a reduction in error connected with estimates of saturated pixels being confused with the sun. The image has a glare in the upper left corner, which causes a misidentification in the area around it as the highest CBH in high-resolution images. This effect is not present in lower resolution images, due to the downsampling and lowpass filtering, resulting in a lower and more correct CBH estimate. 
\begin{figure}[!h] 
    \centering
    \includegraphics[width=3.5in]{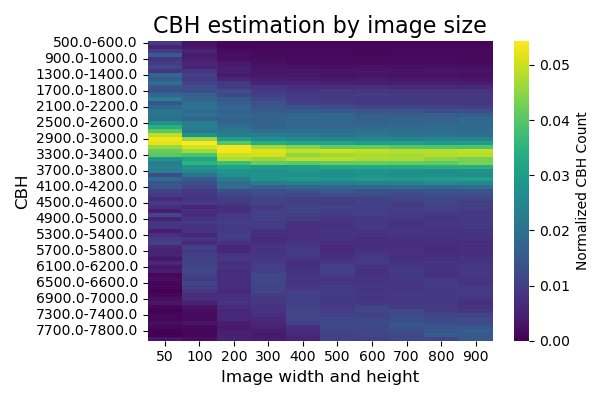} 
    \includegraphics[width=3.5in]{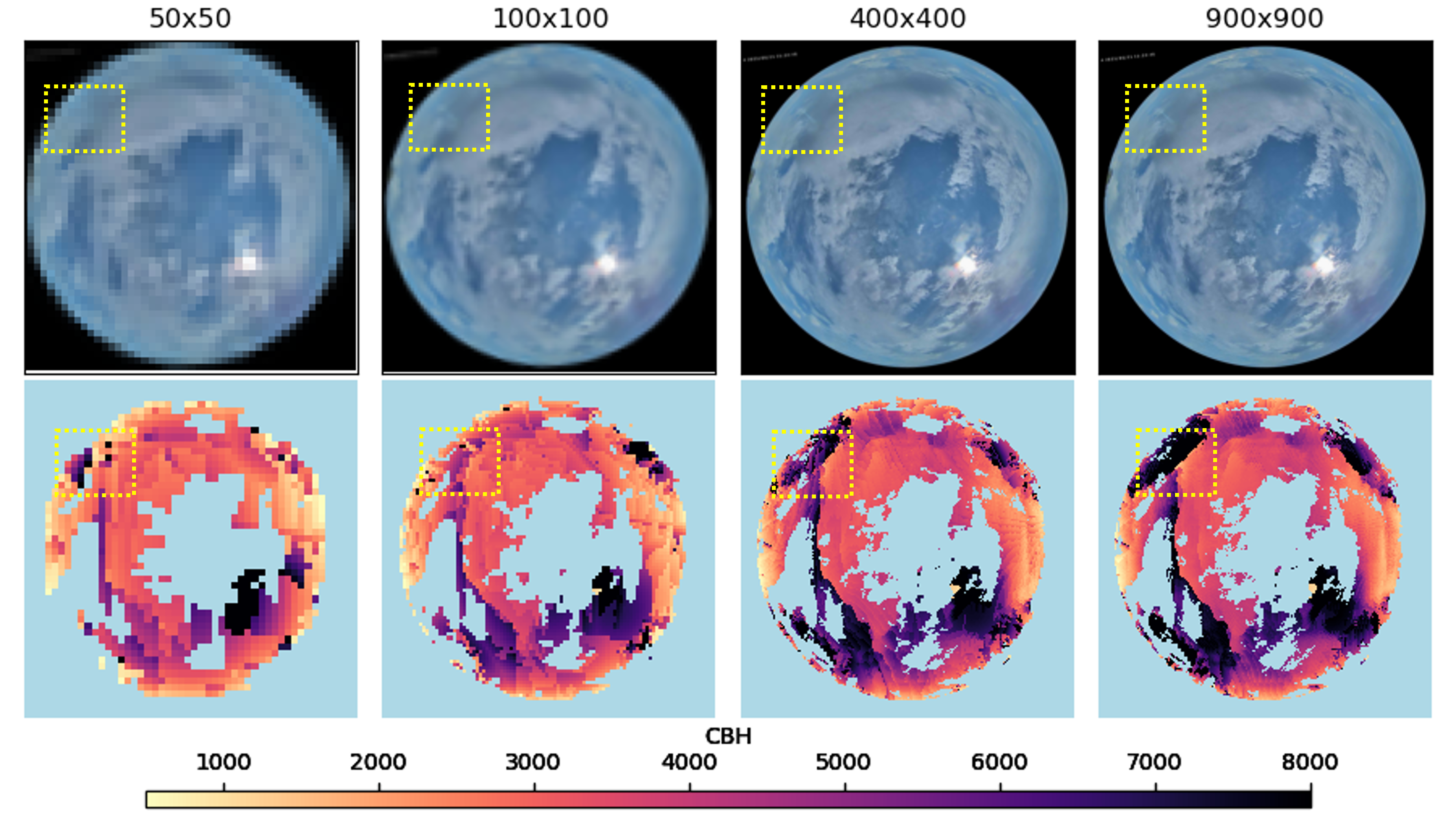} 
    \caption{Distribution and plots of CBH estimates in an image taken on 2023-06-15 13:32:40 CET, relating to image resolution. An area with an image glare is marked in the dashed yellow area, showing how the misidentification of this area reduces with smaller image sizes.}
    \label{appendix:CBH_resolution}
\end{figure}



\clearpage

\section*{Acknowledgment}
The authors acknowledge funding from the Research Council of Norway through the projects KSP-K HYDROSUN (grant number 328640) and KSP-K REHSYS (grant number 344423). The skillful work of Sigurd Brattheim for instrumental maintenance and ensurance of data quality is greatly acknowledged.

/
\bibliographystyle{IEEEtran}

\bibliography{References.bib}

\end{document}